\documentclass[12pt]{article}
\usepackage{geometry}                % See geometry.pdf to learn the layout options. There are lots.
\usepackage{graphicx}
\usepackage{bm}
\usepackage{amssymb,amsmath,amsthm}
\usepackage{mathrsfs} %script fonts
\usepackage{epstopdf}
\def\TODAY{19 August 2008; Revised 4 November 2008}
\title{\bf Transmission probabilities and the Miller--Good transformation}
\author{{\Large Petarpa Boonserm and Matt Visser}\\[5pt]
School of Mathematics, Statistics, and Computer Science\\
Victoria University of Wellington, New Zealand\\[5pt]
{\sf \small \{petarpa.boonserm,matt.visser\}@mcs.vuw.ac.nz}  }
\date{\TODAY;\\  \LaTeX-ed  \today}                                           
\begin{document}
%------------------------------------------------------------------------------------------------------------------------------------------
\maketitle
%------------------------------------------------------------------------------------------------------------------------------------------
% very standard definitions
%------------------------------------------------------------------------------------------------------------------------------------------
\def\d{{\mathrm{d}}}
\newcommand{\scri}{\mathscr{I}}
\newcommand{\sun}{\ensuremath{\odot}}
\def\J{{\mathscr{J}}}
\def\sech{{\mathrm{sech}}}
\newtheorem{theorem}{Theorem}
%--------------------------------------------------------------------------------------------------------------
\begin{abstract}
%--------------------------------------------------------------------------------------------------------------

Transmission through a potential barrier, and the related issue of particle production from a parametric resonance,  are topics of considerable general interest in quantum physics. The authors have developed a rather general bound on quantum transmission probabilities, and recently applied it to bounding the greybody factors of a Schwarzschild black hole. In the current article we take a different tack --- we use the Miller--Good transformation (which maps an initial Schrodinger equation to a final Schrodinger equation for a different potential) to significantly generalize the previous bound.

\vskip 10 pt

Pacs numbers: 03.65.-w, 03.65.Xp, 03.65.Nk, 

\vskip 10 pt

\end{abstract}
%--------------------------------------------------------------------------------------------------------------

\clearpage

%--------------------------------------------------------------------------------------------------------------
\section{Introduction}
%--------------------------------------------------------------------------------------------------------------

Consider the Schrodinger equation,
\begin{equation}
u(x)'' + k(x)^2 \; u(x) = 0,
\label{E:sde}
\end{equation}
where $k(x)^2 = 2m[E-V(x)]/\hbar^2$.  As long as $V(x)$ tends to finite (possibly different) constants $V_{\pm\infty}$ on left and right infinity, then for $E>\max\{V_{+\infty},V_{-\infty}\}$ one can set up a one-dimensional scattering problem in a completely standard manner --- see for example~\cite{dicke-wittke, merzbacher, landau, shankar, capri, messiah, branson-joachim, liboff}.  The scattering problem is completely characterized by the transmission and reflection \emph{amplitudes} ($t$ and $r$), though the most important aspects of the physics can be extracted from the transmission and reflection \emph{probabilities} ($T= |t|^2$ and $R=|r|^2$).
Relatively little work has gone into providing general analytic bounds on the transmission probabilities, (as opposed to approximate estimates), and the only known result as far as we have been able to determine is this:
\begin{theorem}
Consider the Schrodinger equation (\ref{E:sde}). Let $h(x)>0$ be some positive but otherwise arbitrary once-differentiable function. Then the transmission probability is bounded from below by
\begin{equation}
T \geq \sech^2\left\{ \; \int_{-\infty}^{+\infty} {\sqrt{ (h')^2 + (k^2-h^2)^2} \over 2 h} \; \d x \right\}.
\label{E:bd}
\end{equation}
\end{theorem}
\noindent To obtain useful information, one should choose asymptotic conditions on the function $h(x)$ so that the integral converges --- otherwise one obtains the true but trivial result $T\geq \sech^2\infty = 0$. (There is of course a related bound in the reflection probability, $R$, and if one works with the formally equivalent problem of parametric oscillations, a bound on the resulting Bolgoliubov coefficients and particle production.)

This quite remarkable bound was first derived in~\cite{pra}, with further discussion and an alternate proof being provided in~\cite{ann-phys}. These bounds were originally used as a technical step when studying a specific model for sonoluminescence~\cite{sonoluminescence},  and since then have also been used to place limits on particle production in analogue spacetimes~\cite{analogue} and resonant cavities~\cite{cavity}, to investigate qubit master equations~\cite{qubit}, and to motivate further general investigations of one-dimensional scattering theory~\cite{one-dim}. Most recently, these bounds have also been applied to the greybody factors of a Schwarzschild black hole~\cite{greybody}.

A slightly weaker, but much more tractable,  form of the bound can be obtained by applying the triangle inequality.  For $h(x)>0$:
\begin{equation}
T \geq \sech^2\left\{ \; {1\over2} \int_{-\infty}^{+\infty} 
\left[  |\ln(h)' |  + {|k^2-h^2| \over h} \right]\; \d x \right\}.
\label{E:bd0}
\end{equation}
Five important special cases are:  
\begin{itemize}
\item 
If we take $h=k_\infty$, where $k_\infty = \lim_{x\to\pm\infty} k(x)$,  then we have~\cite{pra, ann-phys}
\begin{equation}
T \geq \sech^2\left\{ \; {1\over2k_\infty} \int_{-\infty}^{+\infty} |k_\infty^2-k^2| \; \d x \right\}.
\label{E:bd1}
\end{equation}
\item
If we define $k_{+\infty} = \lim_{x\to+\infty} k(x)\neq k_{-\infty} = \lim_{x\to-\infty} k(x)$,  
and take $h(x)$ to be any function that smoothly and monotonically interpolates between $k_{-\infty}$ and $k_{+\infty}$, then we have
\begin{equation}
T \geq \sech^2\left\{ \;  {1\over2} \left| \ln\left( {k_{+\infty}\over k_{-\infty}} \right) \right| 
+  {1\over2} \int_{-\infty}^{+\infty} {|k^2-h^2| \over h}   \; \d x \right\}.
\label{E:bd2}
\end{equation}
This is already more general than the most closely related  result presented in~\cite{pra,ann-phys}.
\item
If we have a single extremum in $h(x)$ then
\begin{equation}
T \geq \sech^2\left\{ \;  {1\over2} \left| \ln\left( {k_{+\infty} k_{-\infty}\over h_\mathrm{ext}^2} \right) \right| 
+  {1\over2} \int_{-\infty}^{+\infty} {|k^2-h^2| \over h}   \; \d x \right\}.
\label{E:bd3}
\end{equation}
This is already more general than the most closely related  result presented in~\cite{pra,ann-phys}.
\item
If we have a single minimum in $k^2(x)$, and choose $h^2=\max\{k^2,\Delta^2 \}$, assuming $k^2_\mathrm{min}\leq \Delta^2\leq k_{\pm\infty}^2$, (but still permitting $k^2_\mathrm{min} < 0$, so we are allowing for the possibility of a classically forbidden region), then 
\begin{equation}
T \geq \sech^2\left\{ \;  {1\over2} \ln\left({k_{+\infty} k_{-\infty} \over \Delta^2}\right) +  {1\over2\Delta} \int\limits_{\Delta^2>k^2} |\Delta^2 - k^2| \; \d x \right\}.
\label{E:bd4}
\end{equation}
This is already more general than the most closely related  result presented in~\cite{pra,ann-phys}.
\item 
If $k^2(x)$ has a single minimum and  $0 < k^2_\mathrm{min} \leq k_{\pm\infty}^2$, then
\begin{equation}
T \geq \sech^2\left\{ \;  {1\over2} \ln\left( {k_{+\infty} k_{-\infty} \over k_\mathrm{min}^2} \right) \; \right\}.
\label{E:bd5}
\end{equation}
This is the limit of (\ref{E:bd4}) above as $\Delta\to k_\mathrm{min}>0$, and is one of the special cases considered in~\cite{pra}.
\end{itemize}
In the current article we shall not be seeking to \emph{apply} the general bound (\ref{E:bd}), its weakened form (\ref{E:bd0}), or any of its specializations as given in (\ref{E:bd1})--(\ref{E:bd5}) above. Instead we shall be seeking to \emph{extend} and \emph{generalize} the bound to make it more powerful. The tool we shall use to do this is the Miller--Good transformation~\cite{miller-good}.

\section{The Miller--Good transformation}

Consider the Schrodinger equation (\ref{E:sde}), and consider the substitution~\cite{miller-good}
\begin{equation}
u(x) = {1\over \sqrt{X'(x)}} \; U(X(x)).
\label{E:mg}
\end{equation}
We will want $X$ to be our ``new'' position variable, so $X(x)$ has to be an invertible function, which implies (via, for instance, the inverse function theorem) that we  need $\d X/\d x \neq 0$.  In fact, since it is convenient to arrange things so that the variables  $X$ and $x$ both agree as to which direction is left or right, we can without loss of generality  assert  $\d X/\d x > 0$, whence also $\d x/\d X > 0$.

Now compute  (using the notation $U_X= \d U/\d X$):
\begin{equation}
u'(x) = U_X(X) \,\sqrt{X'} - {1\over2} {X''\over (X')^{3/2} } \,U(X),
\end{equation}
and
\begin{equation}
u''(x) = U_{XX}(X) \; (X')^{3/2}  
- {1\over2} {X'''\over (X')^{3/2} } \,U
+{3\over4} {(X'')^2\over(X')^{5/2} } U.
\end{equation}
Insert this into the original Schrodinger equation, $u(x)'' + k(x)^2 u(x) = 0$, to see that
\begin{equation}
 U_{XX} + \left\{ {k^2\over (X')^{2}}  - {1\over2} {X'''\over (X')^{3} }
+{3\over4} {(X'')^2\over(X')^{4} } \right\} U = 0,
\end{equation}
which we can write as
\begin{equation}
U_{XX} + K^2 \; U = 0,
\label{E:sde2}
\end{equation}
with 
\begin{equation}
K^2 = {1\over (X')^2} 
 \left\{ k^2  - {1\over2} {X'''\over X' }
+{3\over4} {(X'')^2\over(X')^{2} } \right\}.
\end{equation}
That is, a Schrodinger equation in terms of $u(x)$ and $k(x)$ has been transformed into a \emph{completely equivalent} Schrodinger equation in terms of $U(X)$ and $K(X)$. 
You can also rewrite this as
\begin{equation}
K^2 = {1\over (X')^2} 
 \left\{ k^2   + \sqrt{X'} \left(1\over \sqrt{X'}\right)'' \right\}.
\end{equation}
The combination
\begin{equation}
 \sqrt{X'} \left(1\over \sqrt{X'}\right)'' =  
- {1\over2} {X'''\over X' } +{3\over4} {(X'')^2\over(X')^{2} }
\label{E:schwartzian-def}
\end{equation}
shows up in numerous \emph{a priori} unrelated branches of physics and 
is sometimes referred to as the ``Schwartzian derivative''.
\begin{itemize}
\item 
As previously mentioned,  to make sure the coordinate transformation $x\leftrightarrow X$ is well defined we want to have $X'(x)>0$, let us call this $j(x) \equiv X'(x)$ with $j(x)>0$.  We can then write
\begin{equation}
K^2 = {1\over j^2} 
 \left\{ k^2  - {1\over2} {j''\over j }
+{3\over4} {(j')^2\over j^{2} } \right\}
\end{equation}
Let us suppose that $\lim_{x\to\pm\infty} j(x) = j_{\pm\infty} \neq 0$; then $K_{\pm\infty} =  k_{\pm\infty}/ j_{\pm\infty}$, so if $k^2(x)$ has nice asymptotic behaviour allowing one to define a scattering problem, then so does $K^2(x)$.

\item
Another possibly more useful substitution (based on what we saw with the Schwartzian derivative) is to set   $J(x)^{-2} \equiv X'(x)$ with $J(x)>0$. We can then write
\begin{equation}
K^2 = {J^4} 
 \left\{ k^2  +{J''\over J }
\right\}
\end{equation}
Let us suppose that $\lim_{x\to\pm\infty} J(x) = J_{\pm\infty} \neq 0$; then $K_{\pm\infty} =  k_{\pm\infty} J_{\pm\infty}^2$, so if $k^2(x)$ has nice asymptotic behaviour allowing one to define a scattering problem, so does $K^2(x)$.
\end{itemize}
These observations about the behaviour at spatial infinity lead immediately and naturally to the result:
\begin{theorem} 
Suppose $j_{\pm\infty} = 1$, \emph{(}equivalently, $J_{\pm\infty} = 1$\emph{)}. Then the ``potentials'' $k^2(x)$ and $K^2(X)$ have the same reflection and transmission amplitudes, and same reflection and transmission probabilities.
\end{theorem}
\noindent
This is automatic since $K_{\pm\infty} = k_{\pm\infty}$, so  equation (\ref{E:sde}) and the transformed equation (\ref{E:sde2}) both have the same asymptotic plane-wave solutions. Furthermore the Miller--Good transformation (\ref{E:mg}) maps any linear combination of solutions of equation (\ref{E:sde}) into the same linear combination of solutions of  the transformed equation (\ref{E:sde2}). QED.
\begin{theorem}
Suppose $j_{\pm\infty} \neq 1$,  \emph{(}equivalently, $J_{\pm\infty} \neq 1$\emph{)}. What is the relation between the reflection and transmission amplitudes, and reflection and transmission probabilities of the two  ``potentials'' $k^2(x)$ and $K^2(X)$? This is also trivial ---  the ``potentials'' $k^2(x)$ and $K^2(X)$ have the same reflection and transmission amplitudes, and same reflection and transmission probabilities.
\end{theorem}
\noindent
The only thing that now changes is that the properly normalized asymptotic states are distinct
\begin{equation}
{\exp(i k_\infty\, x )\over \sqrt{k_\infty} } \leftrightarrow
{\exp(i K_\infty\, x )\over \sqrt{K_\infty} },
 \end{equation}
but map into each other under the Miller--Good transformation. QED.

%----------------------------------------------
\section{Improved general bounds}
%----------------------------------------------

We already know
\begin{equation}
T \geq \sech^2\left\{ \int_{-\infty}^{+\infty}  \vartheta \; \d x \right\}.
\end{equation}
Here $T$ is the transmission probability, and $\vartheta$ is the function
\begin{equation}
\vartheta = {\sqrt{ (h')^2 + [k^2- h^2]^2}\over2 h },
\label{E:b0}
\end{equation}
with $h(x)>0$. But since the scattering problems defined by $k(x)$ and $K(X)$ have the same transmission probabilities, we also have
\begin{equation}
T \geq \sech^2\left\{ \int_{-\infty}^{+\infty}  \tilde \vartheta \; \d X \right\},
\end{equation}
with
\begin{equation}
\d X = X' \; \d x = j \; \d x,
\end{equation}
and 
\begin{eqnarray}
\tilde\vartheta 
&=& 
{\sqrt{ (h_X)^2 + [K^2- h^2]^2}\over2 h } 
\\
&=&
 {1\over2h} \sqrt{ \left({h'\over X'}\right)^2 + \left[ {1\over j^2} 
 \left\{ k^2  - {1\over2} {j''\over j }
+{3\over4} {(j')^2\over j^{2} } \right\} - h^2\right]^2}
\\
&=& {1\over2hj } \sqrt{ (h')^2 + \left[ {1\over j}
 \left\{ k^2  - {1\over2} {j''\over j }
+{3\over4} {(j')^2\over j^{2} } \right\} - j h^2\right]^2}.
\end{eqnarray}
That is: $\forall h(x)>0,\; \forall j(x)>0$ we now have (the first form of) the improved bound
\begin{equation}
T \geq \sech^2\left\{ \int_{-\infty}^{+\infty}   {1\over2h } \sqrt{ (h')^2 + \left[ 
{1\over j}  \left\{ k^2  - {1\over2} {j''\over j }
+{3\over4} {(j')^2\over j^{2} } \right\} - j h^2\right]^2}  \; \d x \right\}.
\label{E:b1}
\end{equation}
Since this new bound contains \emph{two} freely specifiable functions it is definitely stronger than the result we started from,  (\ref{E:bd}). The result is perhaps a little more manageable if we work in terms of $J$ instead of $j$.
We follow the previous logic but now set
\begin{equation}
\d X = X' \; \d x = J^{-2} \; \d x,
\end{equation}
and
\begin{equation}
\tilde\vartheta 
= 
{\sqrt{ (h_X)^2 + [K^2- h^2]^2}\over2 h } 
= 
{1\over2h} \sqrt{ \left({h'\over X'}\right)^2 + \left[ J^4
 \left\{ k^2  + {J''\over J }\right\}
- h^2\right]^2}.
\end{equation}
That is: $\forall h(x)>0,\; \forall J(x)>0$ we have (the second form of) the improved bound
\begin{equation}
T \geq \sech^2\left\{ \int_{-\infty}^{+\infty}   {1\over2h  } \sqrt{ (h')^2 + \left[ 
J^2  \left\{ k^2  + {J''\over J } \right\} - {h^2\over J^2} \right]^2}  \; \d x \right\}.
\label{E:b2}
\end{equation}
A useful further modification is to substitute $h = H J^2$, then
$\forall H(x)>0,\; \forall J(x)>0$ we have (the third form of) the improved bound
\begin{equation}
T \geq \sech^2\left\{ \int_{-\infty}^{\infty}  
 {1\over2H} \sqrt{ \left[H' + 2H \; {J'\over J}\right]^2 + \left[ 
k^2  + {J''\over J } - H^2\right]^2}  \; \d x \right\}.
\label{E:b3}
\end{equation}
Equations (\ref{E:b1}), (\ref{E:b2}), and (\ref{E:b3}), are completely equivalent versions of our new bound.

%----------------------------------------------
\section{Some applications and special cases}
%----------------------------------------------

We can now use these improved general bounds,  (\ref{E:b1}), (\ref{E:b2}), and (\ref{E:b3}), to obtain several more specialized bounds that are applicable in more specific situations.

%----------------------------------------------
\subsection{Schwartzian bound}
%----------------------------------------------

First, take $h=(\mathrm{constant})$ in equation (\ref{E:b2}), then
\begin{equation}
T \geq \sech^2\left\{ {1\over2} \int_{-\infty}^{\infty}   \left|
{J^2\over h}  \left\{ k^2  + {J''\over J } \right\} - {h\over J^2} \right|  \; \d x \right\}.
\end{equation}
In order for this bound to convey nontrivial information we need $\lim_{x\to\pm\infty} J^4 k^2 = h^2$, otherwise the integral diverges and the bound trivializes to $T \geq 0$.  The further specialization of this result reported in~\cite{pra,ann-phys} and equation (\ref{E:bd1}) above corresponds to $J=(\mathrm{constant})= \sqrt{h /k_\infty}$, which clearly is a weaker bound than that reported here. In the present situation we can without loss of generality set $h\to k_\infty$ in which case 
\begin{equation}
T \geq \sech^2\left\{ {1\over2} \int_{-\infty}^{\infty}   \left|
{J^2\over k_\infty}  \left\{ k^2  + {J''\over J } \right\} - {k_\infty\over J^2} \right|  \; \d x \right\}.
\label{E:pre-schwartzian}
\end{equation}
We now need $\lim_{x\to\pm\infty} J = 1$ in order to make the integral converge. 
If $k^2>0$, so that there is no classically forbidden region, then we can choose $J=\sqrt{k_\infty/k}$, in which case
\begin{equation}
T \geq \sech^2\left\{ {1\over2} \int_{-\infty}^{\infty}   \left|
{1\over\sqrt{k}} \left( {1\over\sqrt{k}} \right)'' \right|  \; \d x \right\}.
\label{E:schwartzian}
\end{equation}
This is a particularly elegant bound in terms of the Schwartzian derivative, [equation (\ref{E:schwartzian-def})], which however unfortunately fails if there is a classically forbidden region. This bound is also computationally awkward to evaluate for specific potentials.
Furthermore, in the current context there does not seem to be any efficient or especially edifying way of choosing $J(x)$ in the forbidden region, and while the bound in equation (\ref{E:pre-schwartzian}) is explicit it is not particularly useful.

%----------------------------------------------
\subsection{Low-energy improvement}
%----------------------------------------------

We could alternatively set $H=(\mathrm{constant})$ in equation (\ref{E:b3}), to derive
\begin{equation}
T \geq \sech^2\left\{ \int_{-\infty}^{\infty}  
\sqrt{ \left[ {J'\over J}\right]^2 + {1\over4H^2} \left[ 
k^2  + {J''\over J } - H^2\right]^2}  \; \d x \right\}.
\end{equation}
In order for this bound to convey nontrivial information we need $\lim_{x\to\pm\infty}  k^2 = k_\infty^2 = H^2$,  $\lim_{x\to\pm\infty}  J' = 0$, and  $\lim_{x\to\pm\infty}  J' = 0$.  Otherwise the integral diverges and the bound trivializes to $T \geq 0$.  Thus
\begin{equation}
T \geq \sech^2\left\{ \int_{-\infty}^{\infty}  
\sqrt{ \left[ {J'\over J}\right]^2 + {1\over4k_\infty^2} \left[ 
k^2  + {J''\over J } - k_\infty^2\right]^2}  \; \d x \right\}.
\end{equation}
Again, the further specialization of this result reported in~\cite{pra,ann-phys} and equation (\ref{E:bd1}) above corresponds to $J=(\mathrm{constant})$, which clearly is a weaker bound than that reported here.
To turn this into something a little more explicit, since $J(x)>0$ we can without any loss of generality write
\begin{equation}
J(x) = \exp\left[ \int \chi(x) \;\d x \right],
\end{equation}
where $\chi(x)$ is unconstrained. This permits is to write
\begin{equation}
T \geq \sech^2\left\{ \int_{-\infty}^{\infty}  
\sqrt{ \chi^2 + {1\over4k_\infty^2} \left[ 
k^2  + \chi^2-\chi' - k_\infty^2\right]^2}  \; \d x \right\}.
\end{equation}
Then by the triangle inequality
\begin{equation}
T \geq \sech^2\left\{ \int_{-\infty}^{\infty}  
\left[
 |\chi| + {1\over2k_\infty} \left| 
k^2  + \chi^2-\chi' - k_\infty^2\right|  \right] \; \d x \right\}.
\end{equation}
A further application of the triangle inequality yields
\begin{equation}
T \geq \sech^2\left\{ \int_{-\infty}^{\infty}  
\left[
 |\chi| +  {|\chi'| \over2k_\infty} +  {1\over2k_\infty} \left| 
k^2  + \chi^2- k_\infty^2\right|  \right] \; \d x \right\}.
\label{E:triangle2}
\end{equation}
Now if $k^2 \leq k_\infty^2$, (this is not that rare an occurrence, in a non-relativistic quantum scattering setting, where $k_\infty^2-k^2 = 2mV/\hbar^2$ and we have normalized to $V_\infty=0$, it corresponds to scattering from a potential that is everywhere positive), then we can choose $\chi^2= k_\infty^2-k^2$ so that
\begin{equation}
T \geq \sech^2\left\{ \int_{-\infty}^{\infty}  
\left.\left[
 |\chi| + {1\over2k_\infty} \left| 
\chi'\right|  \right] \; \d x \right\}\right|_{\chi=\sqrt{k_\infty^2-k^2}}.
\end{equation}
Assuming a unique maximum for $\chi$ (again not unreasonable, this corresponds to a single hump potential) this implies
\begin{equation}
T \geq \sech^2\left\{  
{\left.\sqrt{k_\infty^2-k^2}\right|_\mathrm{max}\over k_\infty} +
\int_{-\infty}^{\infty}  
\sqrt{k_\infty^2-k^2} \; \d x \right\}.
\label{E:nnn}
\end{equation}
This is a new and nontrivial bound, which in quantum physics language, where $k^2 = 2m(E-V)/\hbar^2$, corresponds to
\begin{equation}
T \geq \sech^2\left\{  
\sqrt{V_\mathrm{max}\over E} +
\int_{-\infty}^{\infty}  
{\sqrt{2mV}\over\hbar} \; \d x\right\}.
\label{E:nn}
\end{equation}
If under the same hypotheses we choose $\chi=0$, then the bound reported in~\cite{pra,ann-phys} and equation (\ref{E:bd1}) above corresponds to
\begin{equation}
T \geq \sech^2\left\{  
{1\over2\sqrt{E}} \int_{-\infty}^{\infty}  
{\sqrt{2m}\,V\over\hbar} \; \d x\right\}.
\label{E:oo}
\end{equation}
Thus for sufficiently small $E$ the new bound in equation (\ref{E:nn}) is more stringent than the old bound  in equation (\ref{E:oo}) provided
\begin{equation}
\sqrt{V_\mathrm{max}} < {1\over2} \int_{-\infty}^{\infty}  
{\sqrt{2m}\,V\over\hbar} \; \d x.
\end{equation}
Note the long chain of inequalities leading to these results --- this suggests that these final inequalities (\ref{E:nnn}) and (\ref{E:nn}) are not optimal and that one might still be able to strengthen them considerably.

%----------------------------------------------
\subsection{WKB-like bound}
%----------------------------------------------

Another option is to return to equation (\ref{E:triangle2}) and make the choice $\chi^2=\max\{0,-k^2\} = \kappa^2$, so that $\kappa=|k|$ in the classically forbidden region $k^2<0$, while $\kappa=0$ in the classicallty allowed region $k^2>0$. But then  equation (\ref{E:triangle2})  reduces to
\begin{equation}
T \geq \sech^2\left\{ \; \; \int\limits_{k^2<0} \kappa \; \d x + {\kappa_\mathrm{max}\over k_\infty} + {k_\infty\,L\over2} + \int\limits_{k^2>0} {|k_\infty^2-k^2|\over2 k_\infty} 
\; \d x \right\}.
\label{E:WKB-like}
\end{equation}
Key points here are the presence of $ \int\nolimits_{k^2<0} \kappa \; \d x$, the barrier penetration integral that normally shows up in the standard WKB approximation to barrier penetration, $\kappa_\mathrm{max}$ the height of the barrier, and $L$ the width of the barrier.  These is also a contribution from the classically allowed region (as in general there must be, potentials with no classically forbidden region still generically have nontrivial scattering).  Compare this with the standard WKB estimate:
\begin{equation}
T_\mathbf{WKB}  \approx 
\sech^2\left\{ \; \; \int\limits_{k^2<0} \kappa \; \d x + \ln2 \;\right\}.
\end{equation}
This form of the WKB approximation for barrier penetration is derived, for instance, in Bohm's classic textbook~\cite{Bohm}, and can also be found in many other places. Under the usual conditions applying to the WKB approximation for barrier penetration we have $ \int\nolimits_{k^2<0} \kappa \; \d x \gg 1$, in which case one obtains the more well-known version
\begin{equation}
T_\mathbf{WKB}  \approx 
\exp\left\{ \; \; - 2 \int\limits_{k^2<0} \kappa \; \d x \;\right\}.
\end{equation}
The bound in equation (\ref{E:WKB-like}) is the closest we  have so far been able to get to obtaining a rigorous bound that somewhat resembles the standard WKB estimate. Again we do not expect the bound in equation (\ref{E:WKB-like}) to be optimal, and are continuing to search for improvements on this WKB-like bound.

%----------------------------------------------
\subsection{Further transforming the bound}
%----------------------------------------------

In an attempt to strengthen the inequalities (\ref{E:nnn}) and (\ref{E:nn}), we again  use the fact that $J(x)>0$ to (without any loss of generality) write
$J(x) = \exp\left[ \int \chi(x) \;\d x \right]$, 
where $\chi(x)$ is unconstrained. The general bound in equation (\ref{E:b3}) can then be transformed to: For all $H(x)>0$, for all $\chi(x)$:
\begin{equation}
\label{E:xxx}
T \geq \sech^2\left\{ \int_{-\infty}^{\infty}  
 {1\over2} \sqrt{ \left[{H'\over H} + 2 \chi \right]^2 + 
 {\left[ k^2  + \chi^2 + \chi' - H^2\right]^2\over H^2 }  }  \; \d x \right\}.
\end{equation}
This leaves us with considerable freedom. Regardless of the \emph{sign} of $k^2(x)$, we can always choose to enforce $k^2  + \chi^2 - H^2 = 0$, and so eliminate \emph{either} $\chi$ \emph{or} $H$, obtaining
\begin{equation}
T \geq \sech^2\left\{ \int_{-\infty}^{\infty}  
 {1\over2} \sqrt{ \left[{H'\over H} + 2 \sqrt{H^2- k^2} \right]^2 + 
 {\left[ (\sqrt{H^2- k^2})' \right]^2\over H^2 }  }  \; \; \d x \right\},
 \label{E:ppp1}
\end{equation}
(subject to $H(x)>0$ and $H^2(x)-k^2(x)>0$), and
\begin{equation}
T \geq \sech^2\left\{ \int_{-\infty}^{\infty}  
 {1\over2} \sqrt{ \left[{ (\sqrt{\chi^2+k^2})'\over\sqrt{\chi^2+k^2} } 
 + 2 \chi \right]^2 + 
 {  (\chi' )^2\over \chi^2+k^2 }  }  \; \; \d x \right\},
 \label{E:ppp2}
\end{equation}
(subject to $\chi^2(x)+k^2(x)>0$),
respectively.
Finding an explicit bound is now largely a matter of art rather than method. For example if we take
\begin{equation}
H^2= \max\{ k^2, \Delta^2\} \qquad \hbox{or} \qquad \chi^2 = \max\{ 0, \Delta^2-k^2\}
\end{equation}
then from either  equation (\ref{E:ppp1}) or  equation (\ref{E:ppp2}), again under the restriction that we are dealing with a single-hump positive potential,  we obtain
\begin{equation}
T \geq \sech^2\left\{ 
{1\over2} \ln\left({k_{+\infty} k_{-\infty}\over \Delta^2}\right)  
+ {(\sqrt{\Delta^2-k^2})_\mathrm{max}\over\Delta}
+ \int\limits_{\Delta^2>k^2}  \sqrt{\Delta^2-k^2} \; \d x 
\right\}.
\label{E:final}
\end{equation}
Note that $\Delta$ is a free parameter which could in principle be chosen to optimize the bound, however the resulting integral equation is too messy to be of any practical interest. This bound is somewhat similar to that reported in equations (\ref{E:bd4}) and (\ref{E:nnn}), but there are some very real differences.

%----------------------------------------------
\section{Summary and Discussion}
%----------------------------------------------
The bounds presented in this note are generally not ``WKB-like'' --- apart from the one case reported in equation (\ref{E:WKB-like}) there is no need (nor does it seem useful) to separate the region of integration into classically allowed and classically forbidden regions. In fact it is far from clear how closely these bounds might ultimately be related to WKB estimates of the transmission probabilities, and this is an issue to which we hope to return in the future. 

We should mention that if one works with the formally equivalent problem of a parametric oscillator in the time domain then the relevant differential equation is
\begin{equation}
\ddot u(t) + k(t)^2 \; u(t) = 0,
\label{E:po}
\end{equation}
and instead of asking questions about transmission amplitudes and probabilities one is naturally driven to ask formally equivalent questions about Bogoliubov coefficients and particle production. The key translation step is to realize that there is an equivalence~\cite{pra,ann-phys}:
\begin{equation}
T \leftrightarrow {1\over 1+ N};  \qquad  N\leftrightarrow {1-T\over T}.
\end{equation}
This leads to bounds on the number of particles produced that are of the form $N \geq \sinh^2\{ \hbox{(some appropriate integral)} \}$. 

\bigskip\noindent
To be more explicit about this our new improved bound can be written in any of three equivalent forms:
\begin{itemize}
\item 
For all $H(x)>0$, for all $J(x) > 0$, 
\begin{equation}
\label{E:xx}
T \geq \sech^2\left\{ \int_{-\infty}^{\infty}  
 {1\over2H} \sqrt{ \left[H' + 2H \; {J'\over J}\right]^2 + \left[ 
k^2  + {J''\over J } - H^2\right]^2}  \; \d x \right\}.
\end{equation}
\item
For all $h(x)>0$,  for all $J(x) > 0$, 
\begin{equation}
T \geq \sech^2\left\{ \int_{-\infty}^{\infty}   {1\over2h  } \sqrt{ (h')^2 + \left[ 
J^2  \left\{ k^2  + {J''\over J } \right\} - {h^2\over J^2} \right]^2}  \; \d x \right\}.
%\label{E:b2}
\end{equation}
\item
For all $h(x)>0$,  for all $j(x) > 0$, 
\begin{equation}
T \geq \sech^2\left\{ \int_{-\infty}^{+\infty}   {1\over2h } \sqrt{ (h')^2 + \left[ 
{1\over j}  \left\{ k^2  - {1\over2} {j''\over j }
+{3\over4} {(j')^2\over j^{2} } \right\} - j h^2\right]^2}  \; \d x \right\}.
\end{equation}
\end{itemize}
The equivalent statements about particle production are:
\begin{itemize}
\item 
For all $H(t)>0$, for all $J(t) > 0$, 
\begin{equation}
N \leq \sinh^2\left\{ \int_{-\infty}^{\infty}  
 {1\over2H} \sqrt{ \left[H' + 2H \; {J'\over J}\right]^2 + \left[ 
k^2  + {J''\over J } - H^2\right]^2}  \; \d t \right\}.
\label{E:N1}
\end{equation}
\item
For all $h(t)>0$,  for all $J(t) > 0$, 
\begin{equation}
N \leq \sinh^2\left\{ \int_{-\infty}^{\infty}   {1\over2h  } \sqrt{ (h')^2 + \left[ 
J^2  \left\{ k^2  + {J''\over J } \right\} - {h^2\over J^2} \right]^2}  \; \d t \right\}.
\label{E:N2}
\end{equation}
\item
For all $h(t)>0$,  for all $j(t) > 0$, 
\begin{equation}
N \leq \sinh^2\left\{ \int_{-\infty}^{+\infty}   {1\over2h } \sqrt{ (h')^2 + \left[ 
{1\over j}  \left\{ k^2  - {1\over2} {j''\over j }
+{3\over4} {(j')^2\over j^{2} } \right\} - j h^2\right]^2}  \; \d t \right\}.
\label{E:N3}
\end{equation}
\end{itemize}
In closing, we reiterate that these general bounds reported in equations (\ref{E:b1}), (\ref{E:b2}), and (\ref{E:b3}), their specializations in equations (\ref{E:pre-schwartzian}), (\ref{E:schwartzian}), (\ref{E:nnn}), (\ref{E:nn}), (\ref{E:WKB-like}),  and (\ref{E:final}), and the equivalent particle production bounds in equations (\ref{E:N1})--(\ref{E:N3}), are all general purpose tools that are applicable to a wide variety of physical situations~\cite{sonoluminescence, analogue, cavity, qubit, one-dim, greybody}. Furthermore we strongly suspect that further generalizations of these bounds are still possible.

%------------------------------------------------------------------------------------------------------------------------------------------
\section*{Acknowledgments}
This research was supported by the Marsden Fund administered by the Royal Society of New Zealand. PB was additionally supported by a scholarship from the Royal Government of Thailand. 
%------------------------------------------------------------------------------------------------------------------------------------------

%------------------------------------------------------------------------------------------------------------------------------------------

%------------------------------------------------------------------------------------------------------------------------------------------
\end{document}